\documentstyle[epsfig,longtable]{aipproc}

\newcommand{\be}{\begin{equation}}
\newcommand{\ee}{\end{equation}}
\newcommand{\bea}{\begin{eqnarray}}
\newcommand{\eea}{\end{eqnarray}}
\newcommand{\ba}{\begin{array}}
\newcommand{\ea}{\end{array}}
\newcommand{\bm}{\mbox{\boldmath}}

\newcommand{\bfk}{\mbox{\boldmath $k$}}

\newcommand{\qup}{q^\uparrow}

\newcommand{\qdown}{q^\downarrow}
\newcommand{\st}{{\scriptscriptstyle T}}

\begin{document}
\title
{$\Lambda$ and $\bar \Lambda$ polarization as a measurement of distribution 
and fragmentation functions}

\author{\underline{M. Boglione}$^1$, M. Anselmino$^2$, F. Murgia$^3$}
\address{
$^1$Division of Physics, Faculty of Sciences,
    Vrije Universiteit van Amsterdam\\
    De Boelelaan 1081, 1081 HV Amsterdam\\
$^2$Dipartimento di Fisica Teorica, Universit\`a di Torino and \\
    INFN, Sezione di Torino, Via P. Giuria 1, I-10125 Torino, Italy\\
$^3$Istituto Nazionale di Fisica Nucleare, Sezione di Cagliari\\
    and Dipartimento di Fisica, Universit\`a di Cagliari\\
    C.P. 170, I-09042 Monserrato (CA), Italy \\
}

%\lefthead{LEFT head}
%\righthead{RIGHT head}
\maketitle

\begin{abstract}
A combined analysis of the polarization vector of the $\Lambda$ baryons 
produced in DIS processes may give a relevant insight of the hadronization 
process which governs the transition from partons to physical hadrons and 
precise indications on the mechanisms of $spin$ transfer from 
partons to hadrons. We give here a short review of some interesting results.
\end{abstract}

\section*{Introduction}

We consider $\Lambda$ baryon production in polarized Deep Inelastic 
Scattering (DIS) processes, where lepton and proton may or may not be 
polarized.
The idea is in principle quite simple, in fact the 
$\Lambda$ polarization can easily be
measured by looking at the angular distribution of the 
$\Lambda \to p \pi$ decay (in the $\Lambda$ helicity rest frame) and
the fragmenting parton polarization is determined by the elementary
Standard Model interactions, provided one knows the initial parton
spin state. 

Our calculation is performed in the $\ell-p$ center of mass frame:
the lepton moves along the $z$-axis and the proton moves in the opposite 
direction, with four-momenta $\ell$ and $p$ respectively; $xz$ is the 
lepton-hadron production plane. We denote 
by $S_L$ the (longitudinal) nucleon spin oriented along the 
$z$-axis, and by $S_N$ the (normal) spin oriented along the $y$-axis. 
We will only consider longitudinally polarized leptons with
spins $\pm s_L$ which correspond respectively to $\pm$ helicities.

The three components of the baryon polarization vector, $P_i(B)$, 
as measured in its helicity rest frame, are related to the components of the 
helicity density matrix of a spin 1/2 baryon: 
$P_i(B) =$ Tr$[(\sigma^i\rho(B)]$ ($i = x,y,z$).
See Ref. \cite{noi} for details and definitions.

\subsection*{Different Notations}
Before we give the specific expressions for the $\Lambda$ and $\bar \Lambda$ 
polarization vector components, we would like to spend a few words about 
different notations used by different groups to denote distribution and 
fragmentation functions, and how these are related.
For the distribution functions the following identities hold:
\bea
&&q(x) = f_{q/p}(x) = f_1^q(x)\\
&&\Delta q (x) = g_1^q (x) \\
&&\Delta _T q (x) = \delta q = h_1^q(x)\\
&&\Delta^Nf_{q/p^{\uparrow}}(x,{\bfk}_\st) 
= 2\,\frac{\vert \bm k_\st\vert\,\sin\phi}{M} \; 
f_{1T}^{\perp q}(x,\bm k_\st)
\eea
where $\phi$ is the angle of the quark transverse momentum with respect to 
the proton spin. 
See Ref.\cite{BM99} for more details and a pictorial representation of the 
physical content of these functions.  
For the fragmentation functions, the corresponding relations are
\bea
&&D_{h/q}(z) = D_1^q(z) \\
&&\Delta D _{h/q}(z) = G_1^q(z)\\
&&\Delta \! ^N \! D_{h^\uparrow/q}(z,{\bfk}_\st)\,d^2 {\bfk}_ \st = -2\,\frac{\vert 
\bm k_\st\vert \sin\phi}{M_h}
\,H_1^{\perp}(z,\bm k_\st^\prime)\, d^2 {\bfk}_ \st ^\prime 
\label{Delta}
\eea
where, in this case, $\phi$ is the relative azimuthal angle of the outgoing 
hadron momentum (see Ref. \cite{BM99} for details).
In what follows we will make use of the first set of notations.

\section*{$\Lambda$ and $\bar \Lambda$ polarization }
Let's consider a DIS process in which the lepton is not polarized, whereas 
the proton target has positive helicity.
In general, for a spin $1/2$ baryon, we find:
\be
P_z^{(0,+)} = P_z^{(0, -S_L)} =
\frac{\sum_q e_q^2 \Delta q(x) \Delta D_{B/q}(z)} 
{\sum_q e_q^2  q(x)  D_{B/q}(z)}\,. 
%= \frac{\sum_q e_q^2  g_1^q(x)  G_1^q(z)} {\sum_q e_q^2  f_1(x)  D_1(z)}\,, 
\ee
In the case in which the produced hadrons are $\Lambda + \bar \Lambda$, 
and using the assumptions proposed in Ref. \cite{wsd},  
this expression simply becomes:
\be
P_z^{(0,+)} = 
\frac{\Delta Q'(x)}{Q(x)} \,
\frac{\Delta D_{\Lambda/s}(z)}{D_{\Lambda/q}(z)}
%=
%\frac{\Delta Q'(x)}{Q(x)}  
%\frac{ G_1^{s}(z)}{D_1(z)}
\ee
where
\bea
&&D_{\Lambda/u} = D_{\Lambda/d} = D_{\Lambda/s} =
D_{\Lambda/\bar u} = D_{\Lambda/\bar d} = D_{\Lambda/\bar s}
\equiv D_{\Lambda/q} \label{dl} \nonumber \\ &&
\Delta D_{\Lambda/u}(z, Q^2_0) = \Delta D_{\Lambda/d}(z, Q^2_0) = 
N_u \, \Delta D_{\Lambda/s}(z, Q^2_0) 
\eea
are the fragmentation functions of a quark into a $\Lambda$ baryon, 
unpolarized and longitudinally polarized respectively \cite{noi,wsd}. 
The combinations $Q$, $\Delta Q$, $Q'$ and $\Delta Q'$ are defined as follows 
\bea
&&Q \equiv 4(u + \bar u) + (d + \bar d) + (s + \bar s) \nonumber 
\\
&&\Delta Q \equiv 4(\Delta u + \Delta \bar u) + (\Delta d + \Delta \bar d) 
+ (\Delta s + \Delta \bar s) \nonumber
\\
&&Q' \equiv [4(u + \bar u) + (d + \bar d)] \, N_u + (s + \bar s) \nonumber 
\\
&&\Delta Q' \equiv [4(\Delta u + \Delta \bar u) + (\Delta d + \Delta \bar d)]
\, N_u + (\Delta s + \Delta \bar s) 
\eea
For DIS processes where the lepton is longitudinally polarized and the target 
is unpolarized, we have:
\be
P_z^{(+,0)}  = P_z^{(S_L, 0)} 
= \frac{\sum_q e_q^2 q(x){\Delta D_{B/q}(z)}} 
{\sum_q e_q^2  q(x)  D_{B/q}(z)} \hat A _{LL}(y)
%= 
%\frac{\sum_q e_q^2  f_1^q(x)  {G_1^q(z)}} 
%{\sum_q e_q^2  f_1(x)  D_1(z)} { \hat A _{LL}}
\ee
for any spin $1/2$ baryon, whereas for $\Lambda + \bar \Lambda$ production 
we obtain
\be
P_z^{(+,0)} =
\frac{Q'(x)}{Q(x)} 
\frac{{ \Delta D_{\Lambda/s}(z)}}{D_{\Lambda/q}(z)} 
{ \hat A _{LL}(y)}\,,
%=
%\frac{Q'(x)}{Q(x)}  
%\frac{{  G_1^{s}(z)}}{D_1(z)}{ \hat A _{LL}(y)}
\ee
where $\hat A_{LL} (y)$ is the dynamical factor
\be
\hat A_{LL} (y) = \frac 
{d\hat\sigma^{++}_q - d\hat\sigma^{+-}_q} 
{d\hat\sigma^{++}_q + d\hat\sigma^{+-}_q}
= \frac {y(2-y)}{1+(1-y)^2}\,. 
\ee
In the events in which both lepton and target proton are polarized, we have
\be
P_z^{(+,\pm)} = P_z^{(s_L, \mp S_L)} = 
\frac{
\sum_q e_q^2 [q(x)\hat A _{LL}(y) \pm \Delta q(x)]
{ \Delta D_{B/q}(z)}} 
{\sum_q e_q^2 [q(x) \pm \hat A _{LL}(y) \Delta q(x)] D_{B/q}(z)}\,, 
%&=& 
%\frac{\sum_q e_q^2[f_1(x) \hat A _{LL}(y) \pm  g_1^q(x)] 
%{  G_1^q(z)}} 
%{\sum_q e_q^2  [f_1(x) \pm \hat A _{LL}(y) g_1^q(x)]  D_1(z)} 
\ee
and for $\Lambda + \bar \Lambda$ production
\be
P_z^{(+,\pm)} =
\frac{Q'(x) \hat A _{LL}(y) \pm \Delta Q'(x) }
{Q(x) \pm \Delta Q(x)\hat A _{LL}(y)} 
\frac{{ \Delta D_{\Lambda/s}(z)}}{D_{\Lambda/q}(z)}\,.
%=
%\frac{Q'(x) \hat A _{LL}(y) \pm \Delta Q'(x) }
%{Q(x) \pm \Delta Q(x)\hat A _{LL}(y)}  
%\frac{{  G_1^{s}(z)}}{D_1(z)}
\ee

When the final hadrons are produced from {\it transversely} 
polarized protons we find
\be
P_y^{(0, S_N)} = 
\frac{\sum_q e_q^2 {  \Delta _T q(x)}
{ \Delta _T D_{B/q}(z)}} 
{\sum_q e_q^2  q(x)  D_{B/q}(z)} {  \hat D _{NN}(y)} 
%\\ &\propto& 
%\frac{\sum_q e_q^2  {  h_1^q(x)}  {  H_1^{\perp (1)}(z)}} 
%{\sum_q e_q^2  f_1(x)  D_1(z)} {  \hat D _{NN}(y)}
\ee
which, for $\Lambda + \bar \Lambda$ production and under the same assumptions 
(10) for the fragmentation of transversely polarized quarks, reduces to
\be
P_y^{(0, S_N)} = \frac{{ \Delta _T Q'(x)}}{Q(x)} 
\frac{{ \Delta _T D_{\Lambda/s}(z)}}{D_{\Lambda/q}(z)} 
{ \hat D _{NN}(y)}
%\propto
%\frac{{  \Delta _T Q'(x)}}{Q(x)}  
%\frac{{ H_1^{\perp (1) s}(z)}}{D_1(z)}
%{ \hat D _{NN}(y)}
\ee
where
\be
\hat D_{NN} (y) = \frac 
{d\hat\sigma^{\ell \qup \to \ell \qup} -
 d\hat\sigma^{\ell \qup \to \ell \qdown}}
{d\hat\sigma^{\ell \qup \to \ell \qup} +
 d\hat\sigma^{\ell \qup \to \ell \qdown}} =
\frac{2(1-y)}{1+(1-y)^2}
\ee
and
\be
\Delta_T Q' \equiv  
[4(\Delta_T u + \Delta_T \bar u) + 
(\Delta_T d + \Delta_T \bar d)] \, N_u + (\Delta_T s + \Delta_T \bar s)\,.
%[4(h_1^u + h_1^{\bar u}) + 
%(h_1^d +  h_1^{\bar d})] \, N_u + (h_1^ s + h_1^{\bar s})
\ee

Notice that these equations hold within QCD factorization
theorem at leading twist and leading order in the coupling constants;
the intrinsic $\bfk_\st$ of the partons have been integrated over and 
collinear configurations dominate both the distribution and 
the fragmentation functions. Furthermore, for simplicity of notations we 
have not indicated the $Q^2$ scale dependences in $f$ and $D$. 

The measurable components of the $\Lambda$ polarization vector depend on 
different combinations of distribution functions, elementary dynamics and 
fragmentation functions: each of these terms predominantly depends on a 
single variable, respectively $x$, $y$ and $z$, and a careful analysis of 
different situations can yield precious information. 
For example, the longitudinal polarization induced to the baryon by a 
longitudinally polarized nucleon, $P_z^{(0,+)}$, does not depend on the 
elementary dynamics, but only on the quark and spin distribution and 
fragmentation properties (see Eq.(8)). So, neglecting $Q^2$ evolution, 
it does not depend on the variable $y$ but only on $x$ and $z$. 
On the other side, longitudinal polarization induced to the baryon by a 
longitudinally polarized lepton, or by polarized nucleon and lepton at 
the same time, $P_z^{(+,0)}$ and $P_z^{(+,+)}$, depend also on the $y$ 
variable through $\hat A_{LL}(y)$, but in differently weighted 
combinations in the various possible cases.

More specific conclusions can be drawn if one, 
for example, starts by exploiting $P_z^{(+,0)}$ and $P_z^{(0,+)}$ 
to determine $N_u$ and $\Delta D_{\Lambda/s} (z)$. It is then possible to 
apply this knowledge to $P_z^{(+,\pm)}$, which can be used to test the 
validity of the model or suggest its weaknesses and merits.

In the case of {\it transverse} polarization of the proton, if 
$P_y^{(0,S_N)}$ is measured, very important information not only on the 
fragmentation function $\Delta _T D$, but also on the transversity 
distribution function $\Delta _T q(x)$ could be achieved.

Very interesting numerical estimates can be performed by using existing 
models and parameterizations for the distribution and fragmentation functions 
which appear in Eqs~(9),(13),(15) and (18). For details, plots and results 
see Ref. \cite{noi}. 

One last comment before we finish. In the case in which one could use data on 
single cross sections instead of ratios, much more information could be 
gained from this kind of experiments. A very extensive work which exploits 
these new ideas is currently in preparation \cite{sp-fl}.  

\newpage

\subsection*{Conclusions}

The study of the angular distribution of the $\Lambda \to p\pi$
decay allows a simple and direct measurement of the components
of the $\Lambda$ polarization vector. For $\Lambda$'s produced
in the current fragmentation region in DIS processes, the component 
of the polarization vector are related to spin properties of the
quark inside the nucleon, to spin properties of the quark hadronization, 
and to spin dynamics of the elementary interactions.
  
We have discussed all different polarization states of baryons, obtainable
in the fragmentation of a quark in DIS with polarized initial leptons 
and nucleons, showing how they can reveal
different quark features, weighted and shaped by elementary dynamics.

\subsection*{Acknowledgements}
M. Boglione would like to thank the organizers of the workshop for creating 
interesting and stimulating discussion opportunities.

\end{document}